# Co-Design of Free-Space Metasurface Optical Neuromorphic Classifiers for High Performance

François Léonard[†]*, Adam S. Backer[‡], Elliot J. Fuller[†], Corinne Teeter[‡], Craig M. Vineyard[‡]

[†]Sandia National Laboratories, Livermore, CA, 94550, USA

[‡]Sandia National Laboratories, Albuquerque, NM, 87185, USA

*fleonar@sandia.gov

**ABSTRACT.** Classification of features in a scene typically requires conversion of the incoming photonic field into the electronic domain. Recently, an alternative approach has emerged whereby passive structured materials can perform classification tasks by directly using free-space propagation and diffraction of light. In this manuscript, we present a theoretical and computational study of such systems and establish the basic features that govern their performance. We show that system architecture, material structure, and input light field are intertwined and need to be co-designed to maximize classification accuracy. Our simulations show that a single layer metasurface can achieve classification accuracy better than conventional linear classifiers, with an order of magnitude fewer diffractive features than previously reported. For a wavelength λ, single layer metasurfaces of size $100\lambda \times 100\lambda$ with aperture density $\lambda^{-2}$ achieve ~96% testing accuracy on the MNIST dataset, for an optimized distance $\sim 100\lambda$ to the output plane. This is enabled by an intrinsic nonlinearity in photodetection, despite the use of linear optical metamaterials. Furthermore, we find that once the system is optimized, the number of diffractive features is the main determinant of classification performance. The slow asymptotic scaling with the number of apertures suggests a reason why such systems may benefit from multiple layer designs. Finally, we show a trade-off between the number of apertures and fabrication noise.

Classification of objects in a scene is an important task for many applications. The typical approach is to capture the light from the scene with a highly pixelated photodetector which converts the captured photons to the electrical domain. The electronic information is then analyzed by powerful computers using high-end algorithms. This infrastructure imposes serious constrains on power, weight, and throughput. While improvements are possible using neural network algorithms, the approach still requires significant resources. Recently, an approach to circumvent both limitations has been proposed based on passive metamaterials that diffract free-space optical fields(1-8). The free-space propagation allows information processing and transfer at the speed of light until it is captured by a few low-power and fast photodetector pixels. Thus, overall speed and energy efficiency can in principle be significantly improved compared with the conventional optoelectronic approach. Hybrid approaches where free-space optical front ends are integrated with conventional neural networks have also been explored(9, 10).

The free-space diffraction system may also be of interest for accelerating or replacing on-chip computation, an envisioned use for optical neuromorphic computing(11, 12). In this case the system may not directly capture a light field, but artificially inject one based on a problem to be solved. There, the advantage of using programmable device elements is balanced by the need to propagate the optical signals in waveguides which limits the bandwidth and ultimate operating speed in addition to requiring complex fabrication approaches, which impact the scalability due to device footprint. The free-space approach may address some of these issues provided that appropriate system architectures are identified and optimized(10, 13).

The promises of free-space diffractive systems open several interesting questions about the working principles, optimal designs, and ultimate performance. For example, despite the use of linear optical materials, Refs(1, 2) report classification accuracies on the MNIST and Fashion datasets that surpass those of conventional linear classifiers, as well as performance improvements with increasing number of diffractive layers. Understanding the factors that govern the system behavior is not only important to shed light on these results, but also to improve classification performance and reduce the complexity of the system. For instance, having fewer diffractive layers is desirable to increase the amount of transmitted light and reduce problems with alignment, while fewer apertures reduce the fabrication challenges. More generally, more complex problems can be solved for a given system size if system efficiency is optimized.

Here we present extensive simulation results for a *single layer* linear diffractive metasurface with classification accuracy on the MNIST and Fashion-MNIST datasets better than conventional electronic linear classifiers, including neural networks. We show that this is possible using a metasurface with an order of magnitude fewer apertures than previously reported, but that it requires judicious design of the system. In particular, it requires co-design of the structure of the material, the architecture of the system, and the form of the input light field. We show how the system can be mapped into a conventional neural network representation, which allows us to identify an intrinsic nonlinearity due to the light intensity measurement at the output detectors. While the nonlinearity allows classification performance better than conventional linear networks, its presence at the output layer has important implications for system design. In particular, we show that a single metasurface displays slow asymptotic scaling of the testing accuracy with the number of apertures, suggesting an origin for the improvement of performance with multiple layers. We also reveal a trade-off between the number of apertures and the robustness to fabrication noise.

**RESULTS AND DISCUSSION**

The system geometry is illustrated in Fig.1. We consider a single metasurface of size $L \times L$ consisting of $N$ apertures arranged in a square grid (aperture spacing $d$ so that $L = \sqrt{N}d$). Incoming light is transmitted through the metamaterial only through the apertures, which are labelled $k = 1,...,N$ and are located at positions $\vec{r}_k = (x_k, y_k, z = 0)$. The input light field is constructed from the individual images of the MNIST or Fashion-MNIT dataset. Each image consists of $N_{in} \times N_{in} = 28 \times 28$ greyscale pixels which we label with the indices $(m,n)$ and positions $\vec{r}_{mn} = (ma, na, 0)$ where $a$ is the width of the pixel. Throughout this manuscript we set $N_{in}a = L$, i.e. the input light fills the metamaterial fully. The image is converted to a continuous monochromatic light field of wavelength $\lambda$ by distributing the intensity from each pixel $I_{mn}$ over a gaussian of width σ centered on each pixel. Thus the input light intensity from any given pixel $m,n$ at aperture $k$ is $I_{in}^{mn}(\vec{r}_k) = I_{mn} \exp\left(-\frac{|\vec{r}_k - \vec{r}_{mn}|^2}{2\sigma^2}\right)$. Since the light is coherent, the total input intensity at an aperture is constructed by summing the intensities from each pixel $I_{in}(\vec{r}_k) = \sum_{mn} I_{in}^{mn}(\vec{r}_k)$ from which we obtain the total input electric field $E_{in}(\vec{r}_k) = \sqrt{I_{in}(\vec{r}_k)}$.

Each subwavelength aperture diffracts the input light creating an output field according to the Rayleigh-Sommerfeld theory

$$E_{out}^k(\vec{r}) = E_{in}(\vec{r}_k) \frac{z}{|\vec{r}-\vec{r}_k|^2} \left( \frac{1}{2\pi |\vec{r}-\vec{r}_k|} - \frac{i}{\lambda} \right) e^{i\left(\frac{2\pi}{\lambda}|\vec{r}-\vec{r}_k| + \phi_k\right)}. \tag{1}$$

To classify objects from the light field we define $M$ output photodetectors corresponding to $M$ object classes. Typically, $M$ is a small number compared to the number of pixels in a conventional imaging photodetector, leading to significantly lower energy costs when considering the energy required to operate, read-out, and analyze the signals. In this work, we consider photodetectors distributed in a circular pattern on a plane, as illustrated in Fig. 1, with $\vec{r}_p$ denoting the position of detector $p$. The radius of the circle is $R$ and the plane is separated from the material by distance $H$. We chose the digit "0" (item 1 for Fashion) for the center detector and evenly distributed the remaining nine detectors around the circle. (Other configurations are possible(1); we obtained similar results with all 10 detectors distributed in a circle). The light intensity on detector $p$ is calculated from $I_p = \left| \sum_k E_{out}^k(\vec{r}_p) \right|^2$.

Our goal is to optimize the phases $\phi_k$ such that the intensity on detector $p$ is maximal when the incoming light field contains an object corresponding to class $p$. This is done by minimizing the cross-entropy cost function

$$C = -\frac{1}{N} \sum_{images} \log \left( \frac{e^{\tilde{I}_t}}{\sum_{p=1}^{M} e^{\tilde{I}_p}} \right) \tag{1.2}$$

where the sum is over the $N$ training images, and the subscript $t$ indicates the target detector. We also used a normalization of the intensities(2)

$$\tilde{I}_p = f \frac{I_p}{\max\{I\}} \tag{1.3}$$

where $\max\{I\}$ means the maximum value of the intensity on the $M$ detectors; we tested a few values for $f$ and found that $f = 10$ gives the best results in agreement with Ref. (2).

The optimization is accomplished numerically using a gradient descent approach with the Adam algorithm(14). We used a learning rate scheduler where the learning rate was halved if the training accuracy decreased between two successive epochs. The phases were updated at each epoch using the known analytical expressions for $\partial C / \partial \phi_k$. The process was implemented in Fortran on a single 2.7 GHz processor with 64GB of RAM.

*Datasets and training.* The MNIST(15) and Fashion(16) datasets were used in their original format and order. Both contain $M = 10$ classes: for MNIST these correspond to the digits 0 to 9 while for Fashion these are ten different types of clothing. We trained on the first 50,000 images and tested on 10,000 images. The phases were optimized using mini-batches of 20 images, which leads to an improvement of the convergence rate but similar testing accuracies as full-batch training. The inference performance was obtained by calculating the optical intensity on each of the output detectors as the test image light fields went through the trained phase mask. A successful classification was counted when the output light intensity was maximal on the target detector. An example training convergence is shown in the Supplementary Information.

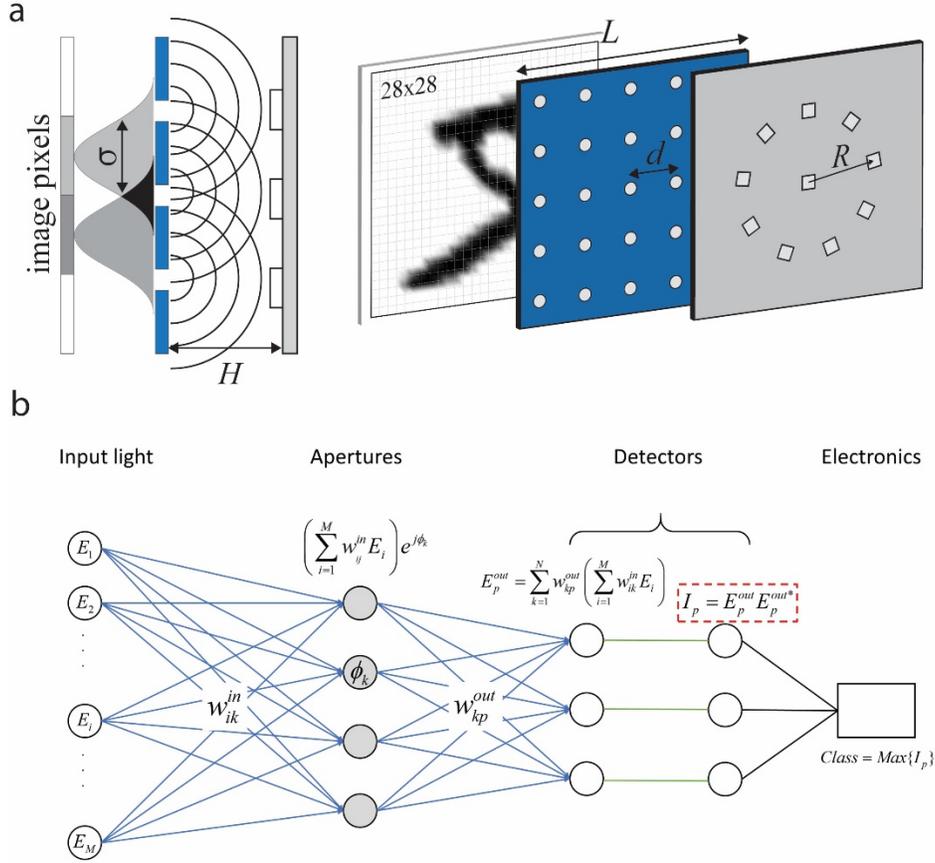

Figure 1: Geometry of the free-space passive classification system. (a) The input light field impinges on a metasurface consisting of subwavelength apertures. The diffracted light is modulated by the metasurface to maximize intensity on photodetectors corresponding to each class of object to be classified. The light from each input pixel is spread over a lateral size σ. The separation between the metamaterial and the detector plane is $H$. The input light field consists of 28x28 pixels, as illustrated with the handwritten number 5. The metamaterial is of side length $L$ and aperture spacing $d$. The detector plane has 10 detectors arranged in a circle of radius $R$. (b) Network representation of the optical system. The input light is represented by the electric field at pixels, with the light from each pixel propagating to the aperture plane. An input pixel $i$ is connected to an aperture $k$ through the complex weight $w_{ik}^{in}$. Each aperture sums the electric fields from each input pixel and emits a spherical wave that propagates to the detector plane.

Each aperture $k$ with phase $\phi_k$ couples to each detector $p$ through the weight $w_{kp}^{out}$. Each detector adds the electric fields from all the apertures and then converts the electric field to an intensity measurement through a nonlinear process shown with the dashed red box. Only a few pixels, apertures, and detectors are shown for simplicity. While each layer is shown as a one-dimensional vector, in reality each layer is two-dimensional.

A conventional neural network representation of the single layer optical system is shown in Fig. 1b. There, each pixel of the input light field is represented by an input node (while displayed as a one-dimensional array here, it is understood that these form a two-dimensional array). Similarly, each aperture is a node that connects to all the input nodes through the light propagation, forming a fully-connected network. The emitted light from each aperture propagates to the detection system, which performs three functions: first, the detectors receive light from all apertures and sum the electric fields; second, the total electric field is converted to light intensity by the photodetectors through the light-matter interaction; third, a small electronic system monitors the detectors and determines which has the largest intensity.

The weights in the network are determined by the properties of the light field and its propagation: $w_{ik}^{in} = E_i \exp\left(-\frac{|\vec{r}_i - \vec{r}_k|^2}{2\sigma^2}\right)$ and $w_{kp}^{out} = \frac{H}{|\vec{r}_k - \vec{r}_p|^2}\left(\frac{1}{2\pi|\vec{r}_k - \vec{r}_p|} - \frac{i}{\lambda}\right)e^{i\left(\frac{2\pi}{\lambda}|\vec{r}_k - \vec{r}_p| + \phi_k\right)}$. Thus, training the phases is equivalent to training the output weights. While the number of weights is the same as a fully connected standard neural network ( $N_{in}^2 \times N$ for input and $N \times M$ for output) in the optical case they are not independent because each weight is determined by the physics of light propagation in the system architecture. For example, $\sigma$ is the only trainable parameter for the input weights, while for the output weights there are $N$ trainable parameters. Additional global training parameters are the positions of the detectors, which in our case means the distance $H$ and radius $R$, and the aperture spacing $d$.

Perhaps the most important point is the presence of a nonlinearity at the output layer. This nonlinearity originates from the photodetector measurement process where the electric field is converted to light intensity: $I \sim |E|^2$. This nonlinearity at the output, which originates from the fundamental properties of photodetection (i.e. the light-matter interaction), is what allows the system to reach high classification accuracy despite the use of linear optical materials, as shown below. In the Supplementary Information, we discuss how this nonlinearity can reproduce nonlinear logic functions.

Having understood the basic system elements, we now explore its fundamental properties through extensive simulations. We present results for a wavelength $\lambda = 10$ $\mu$m but the results apply to any wavelength by rescaling all the length dimensions. Figure 2a shows an example of the classification process whereby the light field for the handwritten digit 0 impinges on the trained metamaterial leading to an output light field on the detector plane. This light field tends to be concentrated near the ten detectors but does not consist of perfect focusing on the target detector. Rather, the system uses variations of the light intensity over small length scales, as shown in the Supplementary Information.

Testing accuracies obtained for our best optimized systems (discussed below) reach 95.89% for MNIST and 83.68% for Fashion (Fig. 2b). The confusion matrices show that the errors during the testing phase are as expected from such datasets: for MNIST most of the errors are in mislabeling the digit 4 as a 9 and vice versa, while for Fashion the images corresponding to the Shirt class are often confused with other garments (Fig. 2c-e). The classification accuracies for the optical system are compared in Fig. 2b to the best conventional linear classifiers implemented on digital computers(16), which reach a maximum accuracy of ~92% for MNIST and ~84% for Fashion for our optimized systems. The optical system clearly surpasses linear classifiers on MNIST while being comparable for Fashion. The well-known result that neural networks require nonlinearities to surpass linear classifiers indicates that the photodetector nonlinearity is key to enabling this performance in the absence of nonlinear optical phenomena in the metamaterial. As discussed further below, this high level of classification accuracy is obtained with only a single layer containing far fewer apertures than previously reported(2) for similar accuracies. We now discuss how the system needs to be co-designed to achieve this high performance and reduce its complexity.

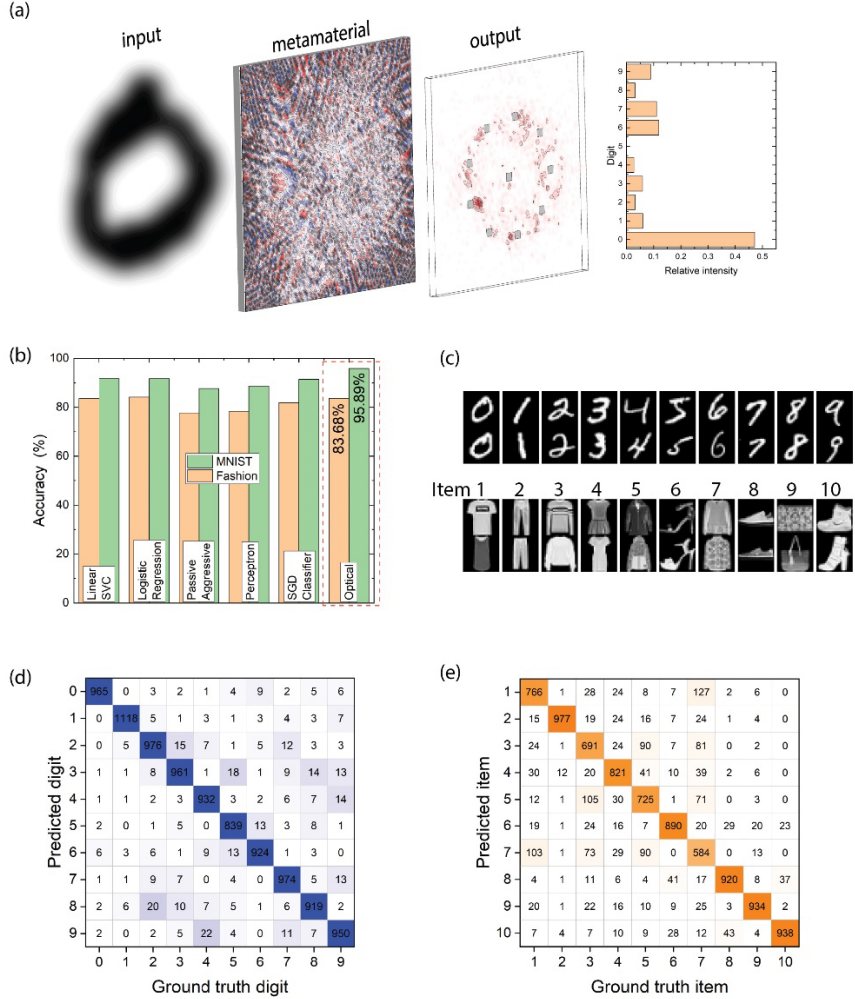

Figure 2: (a) Example of the classification process. The light field for the digit 0 impinges on the metamaterial creating an output field on the detector plane. Here the phases between 0 and $2\pi$ are plotted in a color contour plot on the metamaterial to facilitate visualization. The maximum intensity on one of ten detectors (grey squares) determines the class, as shown in the graph on the right. (b) Test classification accuracy for the MNIST and Fashion datasets for 50,000 training images and 10,000 test images. The results are compared to the best results from conventional linear classifiers from Ref.(16) (c) Example images from the MNIST and Fashion datasets. (c) Confusion matrix for MNIST. (d) Confusion matrix for Fashion. Here $\lambda = 10$ μm, $N = 12544$, $d = 10$ μm, $H = 1300$ μm, $\sigma = 40$ μm.

# IMPACT OF MATERIAL STRUCTURE AND SYSTEM ARCHITECTURE

We begin by studying the impact of the spacing $H$ between the diffraction plane and the detector plane for different metamaterial side lengths $L$. We fix the aperture spacing at $d = 10\mu m$ (the dependence on $d$ is discussed below) and perform a full training of the phases for each $H$. Figure 3a shows the results for four different material lengths $L$ between 140μm, 280 μm, 560 μm, and 1120 μm, which correspond to input pixel sizes of 5μm, 10μm, 20μm, and 40μm, and number of apertures of 196, 784, 3136, and 12544. The detector radii for each are indicated in the figure and were scaled proportionally with $L$. We observe that, for all system sizes, $H$ has a strong impact on performance. There is a lower value $H_{min}$ below which classification is poor and an upper value $H_{max}$ where the classification decreases, with optimized values found between these two limits. As the system size $L$ and the number of apertures increases, the system is able to classify over a broader $H$ range. We find that the system will still classify even for small $H$, and that the larger the system size $L$ the smaller values of $H$ it can tolerate. At these small values of $H$, the denominator in Eq. (1) decreases rapidly with lateral distance and therefore each detector only captures the light from a few apertures. This suggests that in this case (and for this dataset) the classification proceeds through the use of local information from the input light field, and yet relatively high accuracies are reachable.

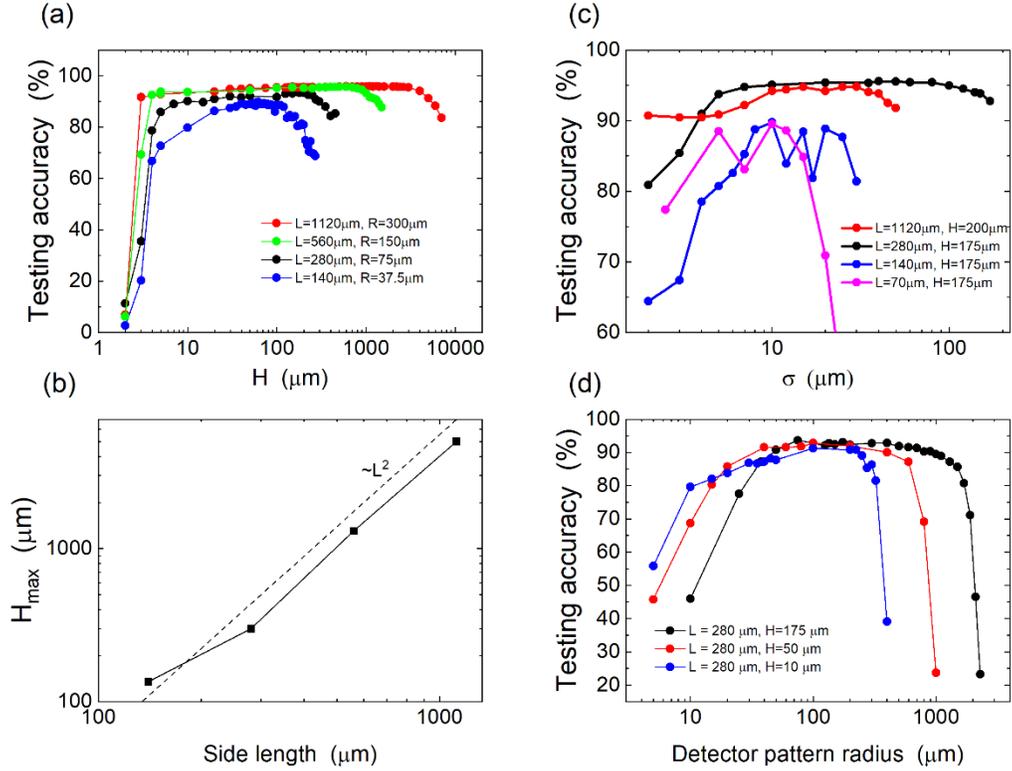

Figure 3: (a) Testing accuracy for the MNIST dataset as a function of the spacing $H$ between the diffraction plane and the detector plane, for several system sizes. Here the aperture spacing is fixed at $d = 10$ μm and the detector pattern radii $R$ are indicated in the legend. (b) Maximum value of $H$ where the testing accuracy begins to decrease for different lateral system sizes $L$. The dashed line indicates a scaling $\sim L^2$. (c) Dependence of testing accuracy on lateral spread of the input pixel for different system sizes $L$ for fixed values of $H$ (given in the legend) and $R$ (given in the legend of panel(a)). (d) Dependence on the detector pattern radius for $L = 280$ μm and fixed values of $H$ (given in the legend) and $R$ (given in the legend of panel(a)).

In contrast to $H_{min}$ the upper limit $H_{max}$ depends strongly on the system size $L$; in Fig. 3b we plot $H_{max}$ defined as the value of $H$ where the testing accuracy decreases below 0.95 of its maximum value. This strong dependence can be understood by considering light focusing by a Fresnel phase mask. For a focal length $f$ the phase distribution in the Fresnel phase mask is given by $\phi(r) = -(2\pi/\lambda)\left(\sqrt{r^2 + f^2} - f\right)$ where $r$ is the radial coordinate. The phase changes by $2\pi$ over a period $l = \lambda\sqrt{1 + 2f/\lambda} \approx \sqrt{2f\lambda}$. Because a few periods are needed to achieve good focus, a system of size $L$ can only accommodate focal lengths up to $f_{max} = L^2/(2n^2\lambda)$ where $n$ is a small integer for the minimum number of periods needed. Figure 4b shows a comparison of this formula for $n = 3$ with the extracted values of $H_{max}$ from the simulations, indicating good agreement with the quadratic scaling.

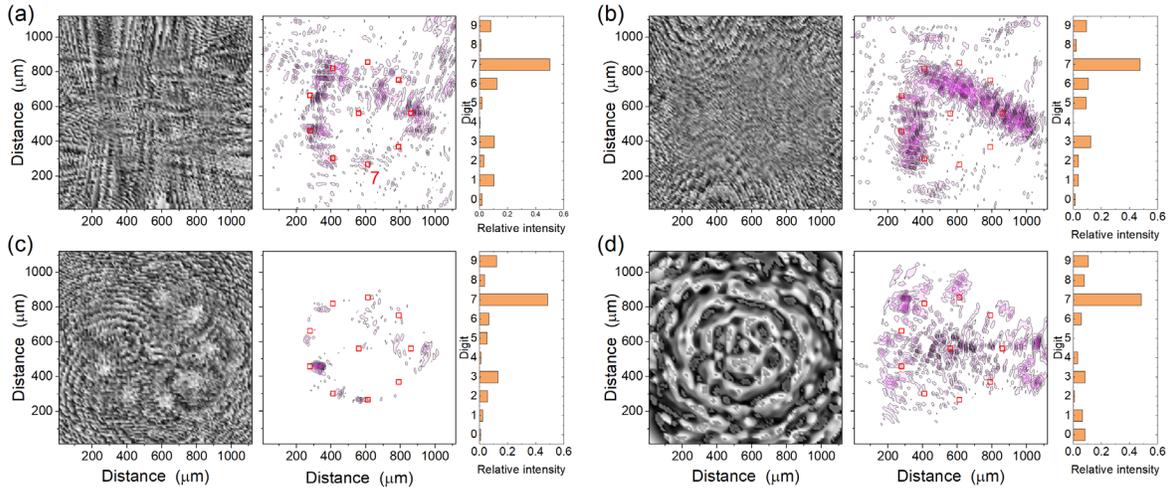

Figure 4: Trained phases for MNIST for various distances between the diffraction and detector planes plotted as gray scale contour plots. Colored images are the relative output light intensities on the detector plane when the number 7 is input through the phase patterns. Relative intensities on each detector are also shown. Here $L = 1120$ µm and $d = 10$ µm. (a) $H = 40$ µm, (b) $H = 300$ µm, (c) $H = 1300$ µm, (d) $H = 7000$ µm. Grey scale is from $-\pi$ to $\pi$. Detector locations are at the center of the red squares, with the number 7 in panel (a) indicating the detector for that digit.

This explanation can be corroborated by considering the trained phases and output light intensities for different $H$. Indeed, Fig. 4 shows the optimized phase distributions for $L = 1120$ μm and $d = 10$ μm for different separations $H$ as well as the output light field for the number 7 (see Supplementary for other examples). There is a clear evolution of the phase pattern from small $H$ to large $H$. At small $H = 40$ μm, the pattern consists mostly of linear features; an outline the handwritten digit 7 appears on the output plane as it is passes through this phase pattern, with small-scale variations that allow classification accuracies ~91%. Increasing $H$ to 300 μm already changes the pattern through the emergence of curved structures, increasing the testing accuracy to ~94%. This becomes a central feature at the optimized spacing $H = 1300$ μm where Fresnel-like circular patterns are clearly seen centered above each detector, and the output light field is much sparser and tends to be focused near the detectors, giving ~96% testing accuracy. Increasing the spacing to 7000 μm reveals the importance of this tenet through the formation of a large central Fresnel-like lens feature which manifests itself as concentric rings of phase shifts about the center of the material. However, at this large $H$ the size of the Fresnel pattern overwhelms the phase distribution, preventing the formation of finer patterns as in Fig. 4c, resulting in a defocused optical field and reduced testing accuracy to ~80%.

Another factor that influences the results is the lateral spread of the light field impinging on the material. So far, we considered a case where $\sigma$ is equal to the pixel size ($\sigma = L/28$). To assess the importance of this parameter we calculated the testing accuracy for fixed $H$ as we varied $\sigma$. (While in principle $\sigma$ is larger than the wavelength or the pixel width, here we also considered small values of $\sigma$ to illustrate the system behavior.) Much like optimizing $H$ was important, we find that $\sigma$ must also be optimized to reach the best performance. Indeed, Fig. 3c shows that the best accuracy is obtained for intermediate values of $\sigma$; this suggests that combining information from adjacent input pixels is useful for classification, but too much overlap leads to blurry input images that are difficult to classify. Furthermore, as the system size $L$ increases the system can accommodate a broader range of σ while maintaining good classification. For the smallest system size with $L = 140$μm there are significant variations of the testing accuracy, with an optimal σ around 10 μm. As the material size increases the optimal σ increases and the maximal efficiency is attained when σ roughly surpasses the input pixel width. Combined with recent work demonstrating that conventional neural networks are able to achieve high classification accuracy when trained on defocused MNIST images(17), this suggests that the features of the dataset are at least as important as the physics of light propagation in determining the optimal σ.

We also explored the role of the detector pattern radius as shown in Fig. 3d. We find that the system classifies in a range of $R$ that depends on the spacing $H$. The reduction in testing accuracy at small $R$ may be due to the difficulty in focusing the light and/or the increased use of local information from the incoming light field, similar to the situation with small $H$. This is more pronounced for larger $H$ because of the radial propagation of the light from the apertures makes it more difficult to focus. We also find an upper cut-off for $R$ beyond which accuracy decreases rapidly. This is most likely due to the difficulty in collecting information from the light field as the detectors extend beyond the edge of the material. We also note that for apertures of finite size we anticipate that $R$ will be more constrained due to the finite angle up to which light can be diffracted. This effect may also impact the overall system performance depending on the importance of capturing long-distance correlations in the input light field.

Having shown that optimizing the detector architecture is important to maximize its performance, we next study the impact of the material structure on the classification performance. Figure 5 shows the dependence of the testing accuracy on the aperture spacing for

*16*

different side lengths $L$. In all cases we observe that the testing accuracy is poor for larger aperture spacings and that it increases significantly as the aperture spacing decreases. While the qualitative trend is similar for all system sizes $L$, there does not appear to be a fundamental quantitative universal dependence on the aperture spacing. On the other hand, the same data plotted in Fig. 5b as a function of the number of apertures $N = L/d$ shows a nearly universal behavior. This indicates that the number of apertures is a fundamental design parameter for the material, determining most of the system performance provided the architecture is optimized. To consider the scaling with $N$ in more detail, we plot in Fig. 5c the testing error vs $1/N$ for MNIST and Fashion. For larger values of $1/N$ there is a power law decrease of the error followed by a significantly slower scaling at smaller $1/N$. The generally slow scaling is reminiscent of the well-known "curse of dimensionality" for conventional neural networks with a single hidden layer and nonlinear activation to the output layer(18). This suggests that the benefit of multilayers/depth for linear optical materials with nonlinear detection could be through the reduction of the total number of apertures needed to attain a certain classification accuracy. While in general it is possible to express a single linear matrix by the product of linear matrices with fewer total elements, it remains to be proven that this can be accomplished with matrices representing the optical transformations necessary for classification.

Taken together, our results indicate an interplay between the number of features in the material and their physical separation. These results are summarized in Fig. 5d where the MNIST testing accuracy is plotted as a function of material side length $L$ for different aperture spacings and number of apertures. Our best accuracy of 95.9% corresponds to a system of 1120 μm side length $L$ with 10 μm aperture spacing and 12,544 apertures. A more compact design with $L =$ 560 μm and $d = 10$ μm achieves nearly the same accuracy (95.71%) with only 3136 apertures.

This can be compared with Ref. (2) where a similar accuracy was obtained using a single layer metasurface containing 40,000 apertures. This can also be compared with conventional neural networks where similar accuracies were initially obtained using >260,000 trainable parameters(15) and since improved to ~100,000 trainable weights(19). The optical system requires more than an order of magnitude fewer training parameters than these approaches. Establishing the origin of this effect requires further studies, but we note that pruning of conventional networks has been demonstrated to reduce the number of needed weights by an order of magnitude or more(20).

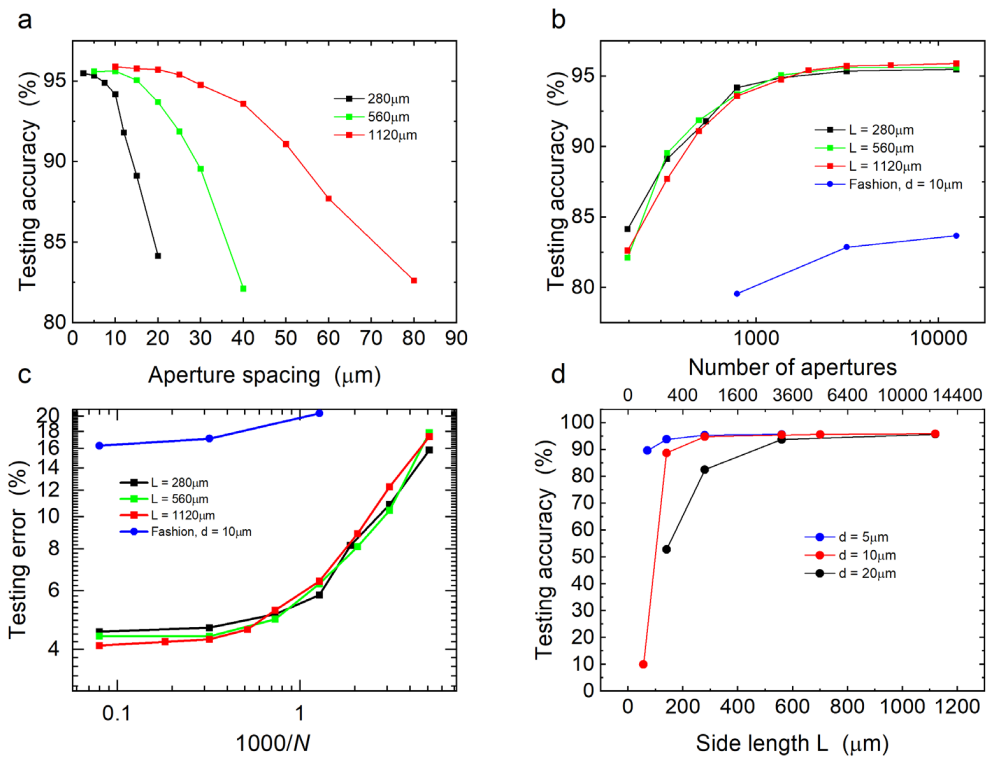

Figure 5: (a) Testing accuracy for MNIST as a function of aperture spacing. (b) Testing accuracy for MNIST and Fashion as a function of the number of apertures. (c) Scaling of MNIST and Fashion testing error with $1/N$. (d) MNIST testing accuracy versus side length.

Transmission of the light through the metasurface and the detection of the output light by the photodetectors ultimately determine the speed, energy cost, and minimum detectable light intensity. The minimum detectable power for a photodetector is given by $P_{min} = NEP/\sqrt{\tau}$ where *NEP* is the noise equivalent power and τ is the integration time. Small area commercial photodiodes(21) achieve NEPs of $5 \times 10^{-15} W/\sqrt{Hz}$ for response times of 1 ns. For integration and recovery times of 5 ns, we calculate $P_{min} = 70$ pW. Thus, classification inference of small light intensities at rates larger than 100 MHz are possible even with significant losses due to transmission through the metamaterial (a simple estimation for apertures of radius *r* would give a transmission coefficient $T = \pi r^2/d^2 = \pi r^2/\lambda^2$, so the detectable *input* power would be about $1/T$ of the above values. In addition, we anticipate that the results obtained here would generalize to other metasurfaces, such as phase arrays, where the optical transmission could be larger.) For typical dark currents of 20 pA at a reverse bias of 5V, the electrical power needed during the integration and recovery times is on the order of 100 pW per class, or about 1 aJ/inference/class. Additional energy is needed to monitor the detectors, which could be accomplished with high speed comparators(22) which have sub-ns response times and require ~100 fJ/conversion. Even with this additional energy cost, the performance compares favorably with conventional neural networks where the best systems currently require ~ 10 mJ/inference, with projections to ~100 μJ/inference(23, 24).

**ROBUSTNESS TO PHASE NOISE**

Fabricating the optical metamaterial requires control over the phases of each aperture. This can be accomplished by controlling the cross-sectional thickness of material inside of each aperture to obtain the desirable phase shift. Fabrication processes can introduce some fluctuations in the desired phases, and here we assess the robustness of the system to these imperfections. We start from a trained phase configuration and add random noise of amplitude α at each aperture, with α taken as a distribution between 0 and $2\pi\alpha$, and calculate the testing accuracy with the new phase profile. We generated 100 noise configurations for each case considered. Figure 6a shows a histogram of the testing accuracy for a system with 784 apertures and $d = 10$ μm, which has a noise-free accuracy of 94.18%. Upon adding 1% noise we observe an overall reduction of the testing accuracy but with the distribution staying relatively narrow. As the noise is increased to 5% and then 10%, the distributions become broader with the 10% noise significantly degrading the system performance. This reduction in performance can be mitigated by going to a system with more apertures, as shown in Fig. 6b. There, the same 10% noise applied to our best performing system ($N = 12{,}544$, $d = 10$ μm) leads to a much narrower distribution. Thus, we find a trade-off between the number of apertures and noise robustness. The noise considered here can be translated to variations on the material thickness $t$ inside each aperture using the relation $\phi = 2\pi n t / \lambda$. Phase variations from 0 to $2\pi$ require thicknesses between 0 and $\lambda / n$; control of the phase noise to within 1% also requires that the thickness be controlled to within $\lambda / 100 n$. In the visible this implies thickness variations less than 1 nm, which is stringent but achievable with modern lithography techniques. Longer wavelengths reduce the fabrication burden at the expense of thicker films.

In addition to variations in the material, other factors could also influence the performance, such as shifts, rotations, and distortions of the input field. These could be addressed by training the system with such light fields, or by working in the Fourier domain using the properties of optical transformations(6). Determining which of these approaches will be most beneficial will require an in-depth study for specific classification tasks.

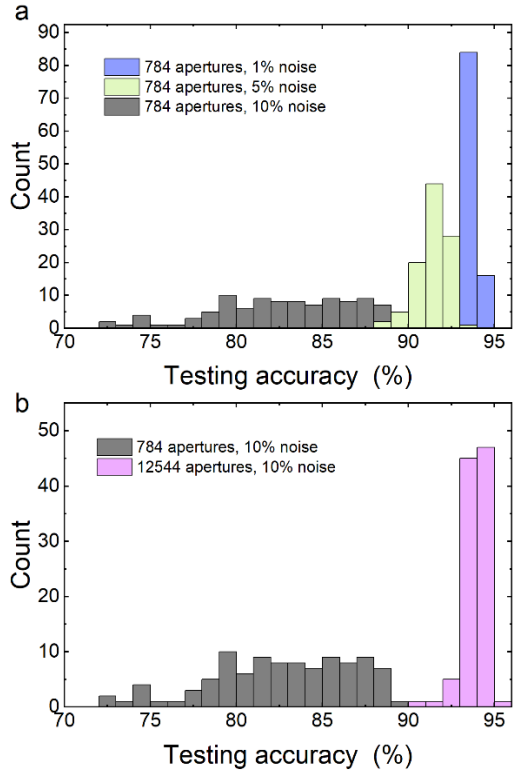

Figure 6: Impact of fabrication phase noise on classification performance. Histogram of 100 noise configurations for (a) three random noise strengths for $N = 784$, $d = 10$ μm, and (b) a fixed noise strength of 10% for $N = 784$ and $N = 12{,}544$, both with $d = 10$ μm.

## CONCLUSIONS

Through detailed numerical simulations of free-space all-optical diffractive classifiers, we demonstrate that co-design can be used to improve performance and reduce system complexity. We find an intricate interplay between the properties of the light field, the material structure, and the overall system architecture. We show that an optimized single linear diffractive layer attains classification performance that surpasses the best linear electronic classifiers with a much reduced number of apertures compared with previous work. We suggest a basis by which these systems can benefit from "depth" through the use of multiple diffractive layers even if only linear optical materials are used. Our results open up a rich set of fundamental research questions. For example, work is needed to understand the set of optical transformations that can be achieved with multiple layers compared to a single layer, and whether these can lead to improved scaling with the number of apertures. We also anticipate that the co-design aspect for multiple layers will be more complex and will necessitate new approaches that can simultaneously train the material and the architecture. Finally, making a connection between the properties of features in an optical scene and the role of the optical classifier in extracting these features is still an open question.

## ACKNOWLEDGEMENTS

Work supported by the DARPA DETECT program and the Laboratory Directed Research and Development Program at Sandia National Laboratories, a multimission laboratory managed and operated by National Technology and Engineering Solutions of Sandia, LLC, a wholly owned subsidiary of Honeywell International, Inc., for the U.S. Department of Energy's National Nuclear Security Administration under Contract No. DE-NA-0003525. The views expressed in the article do not necessarily represent the views of the U.S. Department of Energy or the U.S. Government.

## ASSOCIATED CONTENT

**Supporting information.** This material is available free of charge via the internet at http://pubs.acs.org. Examples of training convergence and output light fields for MNIST; examples of trained phases and output light fields for Fashion; discussion of nonlinear logic functions.

# AUTHOR INFORMATION

## Author contributions

F.L. performed the numerical simulations and analysis. All authors discussed the results and contributed to writing of the manuscript.

## Competing interests

The authors declare no competing financial interest.

# REFERENCES

(1) Lin, X.; Rivenson, Y.; Yardimci, N. T.; Veli, M.; Luo, Y.; Jarrahi, M.; Ozcan, A. All-Optical Machine Learning Using Diffractive Deep Neural Networks. *Science* **2018**, *361*, 1004-1008.
(2) Mengu, D.; Luo, Y.; Rivenson, Y.; Ozcan, A. Analysis of Diffractive Optical Neural Networks and Their Integration with Electronic Neural Networks. *IEEE J. Sel. Top. Quant. Electron.* **2020,** *26*, 1-14.
(3) Wu, Z.; Zhou, M.; Khoram, E.; Liu, B.; Yu, Z. Neuromorphic Metasurface. *Photonics Res.* **2020,** *8*, 46-50.
(4) Li, J.; Mengu, D.; Luo, Y.; Rivenson, Y.; Ozcan, A. Class-Specific Differential Detection in Diffractive Optical Neural Networks Improves Inference Accuracy. *Adv. Photonics* **2019,** *1*, 046001.
(5) Chang, J.; Sitzmann, V.; Dun, X.; Heidrich, W.; Wetzstein, G. Hybrid Optical-Electronic Convolutional Neural Networks with Optimized Diffractive Optics for Image Classification. *Sci. Rep.* **2018,** *8*, 12324.
(6) Yan, T.; Wu, J.; Zhou, T.; Xie, H.; Xu, F.; Fan, J.; Fang, L.; Lin, X.; Dai, Q. Fourier-Space Diffractive Deep Neural Network. *Phys. Rev. Lett.* **2019,** *123*, 023901.
(7) Backer, A. S. Computational Inverse Design for Cascaded Systems of Metasurface Optics. *Opt. Express* **2019,** *27*, 30308-30331.
(8) Ryou, A.; Whitehead, J.; Zhelyeznyakov, M.; Anderson, P.; Keskin, C.; Bajcsy, M.; Majumdar, A. Free-Space Optical Neural Network Based on Thermal Atomic Nonlinearity. *Photonics Res.* **2021,** *9*, B128-B134.
(9) Colburn, S.; Chu, Y.; Shilzerman, E.; Majumdar, A. Optical Frontend for a Convolutional Neural Network. *Appl. Opt.* **2019,** *58*, 3179-3186.
(10) Burgos, C. M. V.; Yang, T.; Zhu, Y.; Vamivakas, A. N. Design Framework for Metasurface Optics-Based Convolutional Neural Networks. *Appl. Opt.* **2021,** *60*, 4356-4365.
(11) Shen, Y.*., et al.* Deep Learning with Coherent Nanophotonic Circuits. *Nat. Photonics* **2017,** *11*, 441-446.
(12) Feldmann, J.*., et al.* Parallel Convolutional Processing Using an Integrated Photonic Tensor Core. *Nature* **2021,** *589*, 52-58.

(13) Tseng, E.; Colburn, S.; Whitehead, J.; Huang, L.; Baek, S.-H.; Majumdar, A.; Heide, F. Neural Nano-Optics for High-Quality Thin Lens Imaging. **2021**, arXiv:1412.6980v9. arXiv.org e-Print archive. https://arxiv.org/abs/2102.11579arXiv:2102.11579 (accessed May 20, 2021).

(14) D. P. Kingma, J. B. Adam: A Method for Stochastic Optimization. **2014**, arXiv:1412.6980v9. arXiv.org e-Print archive. https://arxiv.org/abs/1412.6980v9 (accessed May 20, 2021).

(15) Lecun, Y.; Bottou, L.; Bengio, Y.; Haffner, P. Gradient-Based Learning Applied to Document Recognition. *Proc. IEEE* **1998,** *86*, 2278-2324.

(16) H. Xiao; K. Rasul; Vollgraf, R. Fashion-Mnist: A Novel Image Dataset for Benchmarking Machine Learning Algorithms. **2017**, arXiv:1708.07747v2. arXiv.org e-Print archive. https://arxiv.org/abs/1708.07747 (accessed May 20, 2021).

(17) Zhou, Y.; Song, S.; Cheung, N., On Classification of Distorted Images with Deep Convolutional Neural Networks. In *2017 IEEE International Conference on Acoustics, Speech and Signal Processing (ICASSP)*, 2017; pp 1213-1217.

(18) Cybenko, G. Approximation by Superpositions of a Sigmoidal Function. *Mathematics of Control, Signals and Systems* **1989,** *2*, 303-314.

(19) Yu, D.; Deng, L. Efficient and Effective Algorithms for Training Single-Hidden-Layer Neural Networks. *Pattern Recognition Letters* **2012,** *33*, 554-558.

(20) Babaeizadeh, M.; Smaragdis, P.; Campbell, R. Noiseout: A Simple Way to Prune Neural Networks. **2016**, arXiv:1611.06211. arXiv.org e-Print archive. https://arxiv.org/abs/1611.06211 (accessed May 20, 2021).

(21) https://www.osioptoelectronics.com/Libraries/Datasheets/High-Speed-Silicon-Photodiodes.sflb.ashx (accessed May 20, 2021)

(22) Huang, S.; Diao, S.; Lin, F. An Energy-Efficient High-Speed CMOS Hybrid Comparator with Reduced Delay Time in 40-nm CMOS Process. *Analog Integr. Circuits Signal Process.* **2016,** *89*, 231-238.

(23) Vineyard, C.; Severa, W. M.; Kagie, M.; Scholand, A.; Hays, P. A Resurgence in Neuromorphic Architectures Enabling Remote Sensing Computation. *2019 IEEE Space Computing Conference (SCC)* **2019**, 33-40.

(24) Severa, W. M.; Hill, A. J.; Vineyard, C. M.; Kagie, M. J.; Dellana, R.; Reeder, L.; Wang, F.; Aimone, J. B.; Yanguas-Gil, A. Building a Comprehensive Neuromorphic Platform for Remote Sensing. *Proceedings of the 2019 Government Microcircuit Applications & Critical Technology Conference (GOMACTech)*. **2019**, 1-4.

**TOC image, For Table of Contents Use Only**

Co-Design of Free-Space Metasurface Optical Neuromorphic Classifiers for High Performance

*François Léonard, Adam S. Backer, Elliot J. Fuller, Corinne Teeter, Craig M. Vineyard*

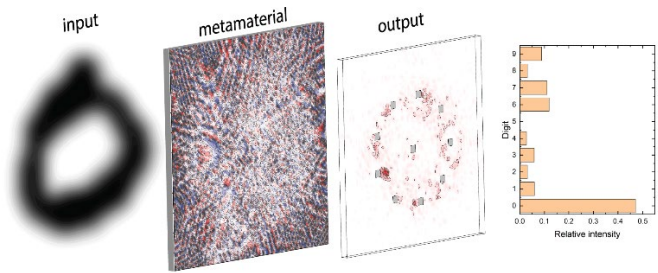

Synopsis: Free-space optical neuromorphic system based on a single layer metamaterial that achieves high accuracy through co-design.

# Supporting Information

Co-Design of Free-Space Metasurface Optical Neuromorphic Classifiers for High Performance

*François Léonard[†]\*, Adam S. Backer[‡], Elliot J. Fuller[†], Corinne Teeter[‡], Craig M. Vineyard[‡]*

[†]Sandia National Laboratories, Livermore, CA, USA

[‡]Sandia National Laboratories, Albuquerque, NM, USA

\*fleonar@sandia.gov

Pages S1-S7.

Figures S1-S5.

Example training convergence for MNIST

Figure S1 shows an example of the training convergence on the MNIST dataset. The accuracy increases quickly in the first few iterations. Jumps in accuracy arise due to the learning rate scheduler where the leaning rate is halved.

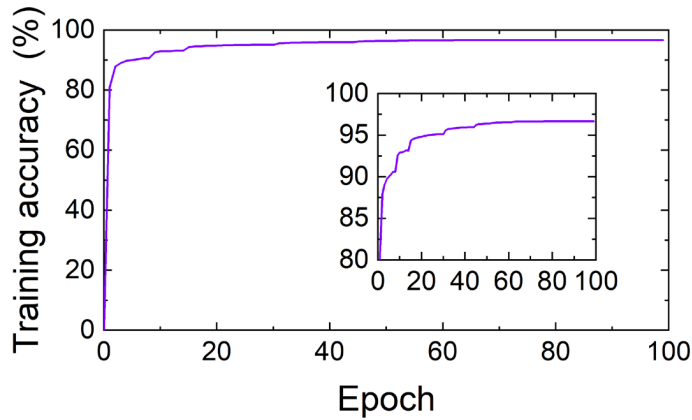

Figure S1. Example of the training convergence for a system with 12,544 apertures training on 50,000 MNIST images. The impact of the learning rate scheduler can be seen in the inset where each jump in accuracy corresponds to halving the learning rate.

Example output light fields for MNIST

Figure S2 shows an example of an optimized phase distribution for an array of 12,544 apertures and the diffracted light field from two digits of the MNIST dataset. The output light faintly traces out the shape of the input digit, with complex detailed variations on shorter length scales. These variations are used to minimize the intensity on the non-target detectors. Notably, the phase mask is able to significantly re-direct the light to the target detectors: indeed, for both numbers the target detectors are located away from regions where the input light intensity is high. In the bottom panels we plot the relative intensity on the detectors (i.e. normalized by the total intensity on the ten detectors) indicating that a clear discrimination can be made. Note that the cost function used above only depends on the light intensity on the ten detectors and therefore the system is not penalized for light outside of the detectors. The cost function could be modified to add this feature, but the need to calculate the light field at a large number of grid points would significantly slow down training. In addition, it has been demonstrated that this approach decreases the classification accuracy[1].

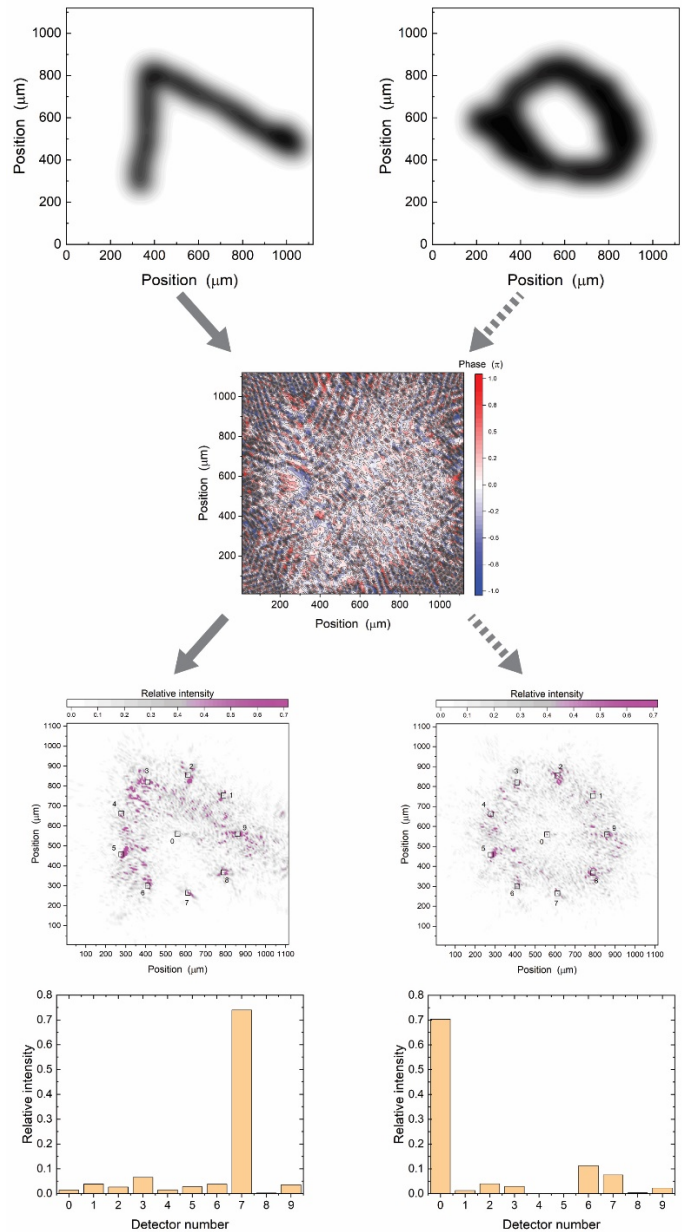

Figure S2: Example of optimized phase configuration for classification of MNIST digits and resulting output light fields. The input light field at the top consists of a handwritten digit. When the light diffracts through the phase mask it is modified to maximize the light intensity on the detector corresponding to that digit. The same phase mask scatters the light from a different digit to a different spatial location. Here the phase mask contains 12,544 apertures, $d = 10\mu m$, and $H = 300um$.

To further illustrate how the system uses small scale variations in the light intensity to achieve classification, we show in Fig. S3 a 100μmx100μm zoom-in of the light field around each detector when the number "0" impinges on the system. Even though at first glance a large intensity spot is present near detector 5, upon closer examination the light intensity is found to be negligible at the detector itself. In fact, only detector 0 has significant intensity.

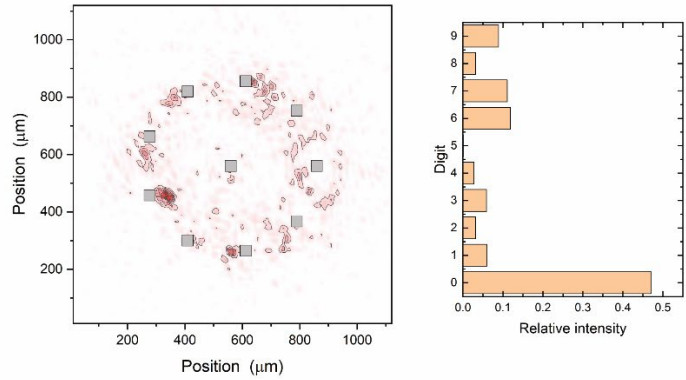
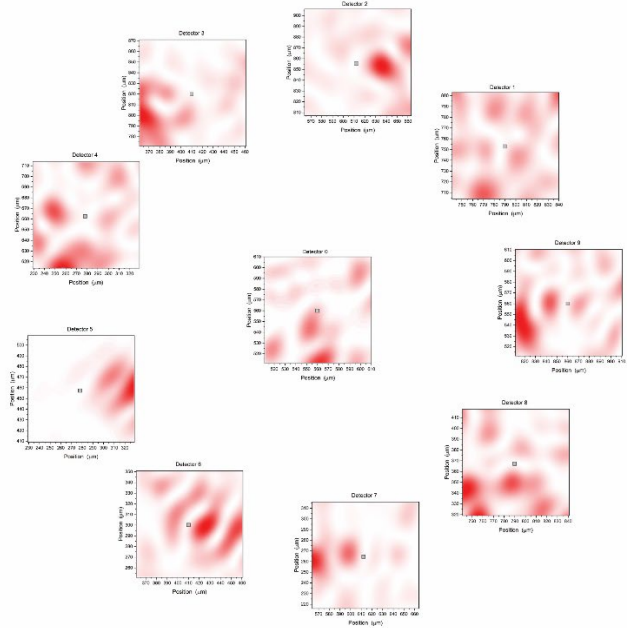

Figure S3: Detailed consideration of output light fields. The input light field consists of the handwritten digit "0". Zoomed-in regions of size 100μm×100μm around each detector show the output light field. Here the phase mask contains 12,544 apertures, $d = 10$μm, and $H = 1300$μm.

Example optimized phases for Fashion

Figure S4 shows an example of a trained phase distribution for the Fashion dataset. Clear circular pattern are observed above each detector. We also show a T-Shirt light field and the output light field when it propagates through the trained phase mask.

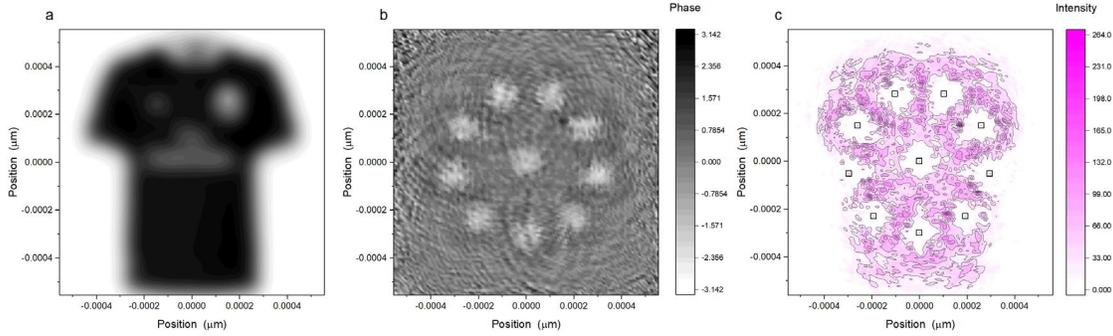

Figure S4: (a) Light field on the material for a Fashion item. (b) Trained phase distribution for H = 1300 μm, d = 10 μm, N = 12,544, σ = 40 μm. (c) Light intensity at the detector plane with the locations of the detectors indicated by the squares.

Nonlinear logic functions

An important property of nonlinear electronic neural networks is their ability to approximate any function, in contrast to linear networks. The canonical example is the XOR function. Here we show that the optical system is able to reproduce this function. We consider a simplified situation with two input pixels of intensities $I_1$ and $I_2$, together with $\sigma = 0$. The optical system consists of two apertures each aligned with an input pixel, and a single detector located between the two apertures, see Fig. S5.

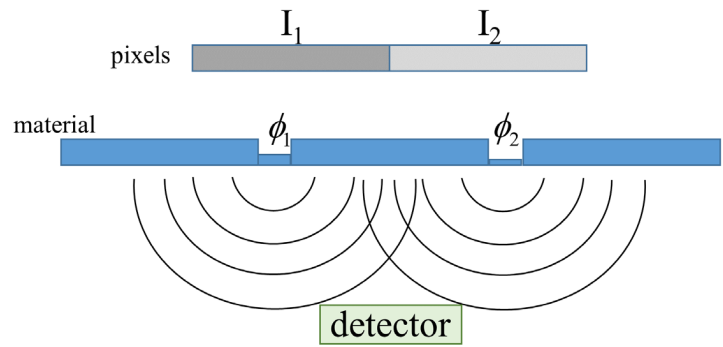

Figure S5: Model optical system for logical functions.

The light intensity on the detector is given by

$$I_{det} = gI_1 + gI_2 + 2g\sqrt{I_1}\sqrt{I_2}\cos(\phi_1 - \phi_2) \tag{S1}$$

where $\phi_1$ and $\phi_2$ are the aperture phases, and $g$ is a constant that depends on the aperture spacing and distance to the detector. By making the choice $\phi_1 - \phi_2 = \pi$ we obtain $I_{det} = 0$ when $I_1 = I_2 = 0$ or $I_1 = I_2$ and $I_{det} = gI$ when $I_1 = I_2 = I$, which is the truth table for the XOR function. This demonstrates that the optical system is able to produce some nonlinear logic functions that linear systems cannot. It is also possible to construct the OR function with $\phi_1 - \phi_2 = 2\pi/3$; however the NOR, AND, and NAND gates are problematic because it is not possible to have zero intensity on the detector when only one aperture is illuminated. For the AND gate this can be overcome by setting $\phi_1 - \phi_2 = 0$ and defining a threshold intensity below which the output is considered zero. Unfortunately, this does not work for the NAND or NOR gates. Since these two gates are universal logic gates from which other logic functions can be built, it suggests that the simple system constructed here has some, but limited, nonlinear capabilities. Note that adding more layers does not solve the fundamental problem that prevents realization of these logic gates.

**References**

1.      Mengu, D.; Luo, Y.; Rivenson, Y.; Ozcan, A., Analysis of Diffractive Optical Neural Networks and Their Integration with Electronic Neural Networks. *IEEE J. Sel. Top. Quant. Electron.* **2020,** *26*, 1-14.